\documentclass[showpacs,showkeys,twocolumn]{revtex4}
\usepackage{amsmath,amsfonts,amssymb}
\usepackage{graphicx}

\def\a{\alpha}
\def\b{\beta}
\def\g{\gamma}

\def\d{\delta}

\def\ve{\varepsilon}

\def\sq2{\sqrt{\frac{\varepsilon_0}{\mu_0}} }  

\begin{document}
\title{Relativistic analysis of magnetoelectric crystals:\\
  extracting a new 4-dimensional $P$ odd and $T$ odd pseudoscalar from
  Cr$_2$O$_3$ data}
\author{Friedrich W. Hehl}
\email{hehl@thp.uni-koeln.de}
\altaffiliation[Also at: ]{Dept. Physics Astron., Univ. of
Missouri-Columbia, Columbia, MO 65211, USA}
\affiliation{Institute for Theoretical Physics, University of Cologne, 
50923 K\"oln, Germany}

\author{Yuri N. Obukhov}
\email{yo@thp.uni-koeln.de}
\altaffiliation[Also at: ]{Department of Theoretical Physics, Moscow 
State University, 117234 Moscow, Russia}
\affiliation{Institute for Theoretical Physics, University of Cologne, 
50923 K\"oln, Germany}

\author{Jean-Pierre Rivera}
\email{Jean-Pierre.Rivera@chiam.unige.ch}
\affiliation{Department of Inorganic, Analytical and
  Applied Chemistry, University of Geneva, Sciences II, 30 quai E.\
  Ansermet, CH-1211 Geneva 4, Switzerland}

\author{Hans Schmid} 
\email{Hans.Schmid@chiam.unige.ch}
\affiliation{Department of Inorganic, Analytical and
  Applied Chemistry, University of Geneva, Sciences II, 30 quai E.\
  Ansermet, CH-1211 Geneva 4, Switzerland}

\begin{abstract}
  Earlier, the linear magnetoelectric effect of chromium sesquioxide
  Cr$_2$O$_3$ has been determined experimentally as a function of
  temperature. One measures the electric field-induced magnetization
  on Cr$_2$O$_3$ crystals or the magnetic field-induced polarization.
  {}From the magnetoelectric moduli of Cr$_2$O$_3$ we extract a
  4-dimensional relativistic invariant pseudoscalar $\widetilde{\a}$.
  It is temperature dependent and of the order of $\sim 10^{-4}\,Y_0$,
  with $Y_0$ as vacuum admittance. We show that the new pseudoscalar
  $\widetilde{\a}$ is odd under parity transformation and odd under
  time inversion.  Moreover, $\widetilde{\a}$ is for Cr$_2$O$_3$ what
  Tellegen's {\it gyrator} is for two port theory, the {\it axion}
  field for axion electrodynamics, and the PEMC (perfect
  electromagnetic conductor) for electrical engineering.
\end{abstract}

\pacs{75.50.Ee, 03.50.De, 46.05.+b, 14.80.Mz}
\keywords{Electrodynamics; Relativity; Constitutive law; 
Magnetoelectric media; Chromium oxide Cr$_2$O$_3$; Broken P and
T invariance; Gyrator; PEMC; Axion electrodynamics}

\date{03 Aug 2007, {\it file PseudoScalar7.tex}}
\maketitle
\section{Introduction}

Our paper addresses the magnetoelectric (ME) effect. This effect has
been established since the 1960's in Cr$_2$O$_3$ crystals (see the
reviews by O'Dell \cite{O'Dell} and, more recently, by Fiebig
\cite{Fiebig2}).  The ME effect, in linear approximation, is described
by magnetoelectric susceptibilities or moduli that have been measured
by different groups.

It has been a long-standing discussion whether these moduli fulfill a
certain condition, as predicted by Post in 1962 \cite{Post}. This
condition was dubbed {\em Post constraint} by Lakhtakia, see, e.g.,
\cite{Akhlesh1},\cite{Akhlesh2}. Numerous arguments against the
validity of the Post constraint were put forward, some of them are
mentioned in \cite{Postconstraint} and \cite{SihTre}, e.g.. However,
in the end, we must turn to the experiments and their proper
evaluation.

Following Post \cite{Post}, we provide here a relativistic invariant
formalism of the electrodynamics of moving media. The violation of the
Post constraint is measured by a {\em pseudoscalar\/} modulus, which
has not been determined so far. If this pseudoscalar vanishes, the
Post constraint is fulfilled, otherwise it is violated. On the basis
of experimental data, we determine this pseudocsalar (or axion) piece
of the magenetoelectric moduli and find it {\em non\/}-vanishing.
Therefore the Post constraint cannot be uphold as a general valid
relation.

However, our paper has also an interdisciplinary purpose. Our result
of the non-vanishing pseudoscalar provides a physical structure that
also shows up in the theory of electric networks, more exactly in the
theory of two ports, as Tellegen's gyrator \cite{Tellegen1948}, in
electrical engineering as perfect electromagnetic conductor (PEMC)
\cite{LindSihv2004a}, and in elementary particle physics as the
hypothetical axion field \cite{Ni},\cite{Wilczek87}. These
interrelationships support each other. Since the axion in elementary
physics is the only left hypothetical object in this context, our
results make also the existence of the axion particle more likely.

  In Sec.2, we give a short description of the ME effect. In Sec.3, a
  4-dimensional electrodynamic framework for moving media is built up
  and the electromagnetic constitutive tensor introduced for local and
  linear media. In Sec.4 we discuss Dzyaloshinskii's theory \cite{Dz1}
  of Cr$_2$O$_3$ and extract therefrom the mentioned pseudoscalar. In
  Sec.5 we finally determine the pseudoscalar for the first time and
  discuss some of its properties in Sec.6. Then, in Sec.7, we turn to
  the interdisciplinary part and discuss the existence of the
  pseudoscalar for network theory, electrical engineering, and
  elementary particle physics. In the concluding section, we collect
  our results.

\section{Magnetoelectric effect}

In classical electrodynamics for a local linear medium, which is at rest in
the reference frame considered, the constitutive law reads
$\mathbf{D}=\ve\varepsilon_0\mathbf{E}$ and
$\mathbf{H}=\mathbf{B}/(\mu\mu_0)$. Here $\ve_0$ is the electric
constant (permittivity of free space) and $\mu_0$ the magnetic
constant (permeability of free space), whereas $\ve$ and $\mu$ are the
(relative) permittivity and permeability, respectively, of the medium
under consideration. Furthermore, the admittance of free space is
$Y_0=1/\Omega_0=\sqrt{\ve_0/\mu_0}$, with $\Omega_0$ as vacuum
impedance, and the speed of light $c=1/\sqrt{\ve_0\mu_0}$.

If an external $\mathbf{B}$ field in some suitable medium induces an
electric excitation $\mathbf{D}$ and an external $\mathbf{E}$ field a
magnetic excitation $\mathbf{H}$, the constitutive law mentioned has
to be extended by so-called magnetoelectric pieces, see O'Dell
\cite{O'Dell}.  The general {\it local} and {\it linear} constitutive
law, if the medium is anisotropic, reads
\begin{eqnarray}
D&=&\hspace{7pt}(\ve)\ve_0\,E+\hspace{11pt}(\a_1)Y_0\,B\,,\label{DEB}\\
H&=&(\a_2)Y_0\,E+(\mu^{-1})\mu_0^{-1}\,B\,.\label{HEB}
\end{eqnarray}
We have to read (\ref{DEB}) and (\ref{HEB}) as tensor equations, with
$(\ve)$, $(\mu^{-1})$, $(\a_1)$, and $(\a_2)$ as dimensionless
$3\times 3$ matrices.  Hence we expect 36 permittivity, permeability,
and magnetoelectric moduli in general.  The constants $\ve_0$, $Y_0$,
and $\mu_0$ are required for dimensional consistency.

The existence of nonvanishing $(\a_1)$ and $(\a_2)$ matrices was
foreseen by Landau-Lifshitz \cite{LL} for certain magnetic crystals and
proposed by Dzyaloshinskii \cite{Dz1} specifically for the
antiferromagnet Cr$_2$O$_3$.  Astrov \cite{Astrov} (for an electric
field) and Rado \& Folen \cite{RadoFolen62} (for a magnetic field)
confirmed this theory experimentally for Cr$_2$O$_3$ crystals. For
reviews, see \cite{A} --- there other magnetoelectric crystals are
listed, too --- and \cite{Fiebig2}.

\section{Four-dimensional Maxwellian framework} 

In order to extend the formalism to systems {\it moving} in the
reference frame considered, but also in order to recognize the
relativistic covariant structures of (\ref{DEB}) and (\ref{HEB}), we
have to go over to a four-dimensional (4D) formalism. We collect $D$
and $H$ in the 4D excitation tensor density
$\frak{G}^{\mu\nu}(D,H)=-\frak{G}^{\nu\mu}$ and $E$ and $B$ in the 4D
field strength tensor $F_{\mu\nu}(E,B)=-F_{\nu\mu}$, with
$\mu,\nu,\dots=0,1,2,3$ and coordinates $(x^0=t,x^1,x^2,x^3)$, see
Post \cite{Post}. Then the Maxwell equations read
\begin{equation}\label{Max}
  \partial_\nu\frak{G}^{\mu\nu}=\frak{J}^\mu\,,\quad \partial_\mu
  F_{\nu\lambda}+\partial_\nu F_{\lambda\mu}+ \partial_\lambda
  F_{\mu\nu}=0\,,
\end{equation}
with $\frak{J}^\mu(\rho,j)$ as 4-current. This is the ``premetric''
form of Maxwell's equations. They are covariant under general
coordinate transformations and do not depend on the metric of
spacetime, that is, they are valid in this form in special and in
general relativity alike.

In order to complete the Maxwell equations (\ref{Max}) to a predictive
physical system, we have to specify a constitutive law linking
$\frak{G}^{\mu\nu}$ to $F_{\mu\nu}$. In vacuum, here now eventually
the metric $g_{\mu\nu}$ of spacetime enters with signature $(-++\,+)$,
we have
 \begin{equation}\label{strel} \frak{G}^{\lambda\nu}
   = Y_0\sqrt{-g}g^{\lambda\a}g^{\nu\b}F_{\a\b} =
   Y_0\sqrt{-g}F^{\lambda\nu}\,,
\end{equation}
with $g:=\det g_{\rho\sigma}\ne 0$. {}From the covariant components of
the metric $g_{\mu\nu}$, its contravariant components $g^{\lambda\nu}$
can be determined via $g_{\mu\lambda}\,g^{\lambda\nu}=
\delta_{\mu}^\nu$.

For magnetoelectric media that are local and linear, we assume,
following Tamm \cite{Tamm} and Post \cite{Post}, the constitutive law
\begin{equation}\label{constit1}
  \frak{G}^{\lambda\nu}=\frac
  12\,\chi^{\lambda\nu\sigma\kappa}F_{\sigma\kappa}\,,
\end{equation}
where $\chi^{\lambda\nu\sigma\kappa}$ is a {\it constitutive tensor
  density} of rank 4 and weight $+1$, with the dimension
$[\chi]=1/resistance$. Since both $ \frak{G}^{\lambda\nu}$ and
$F_{\sigma\kappa}$ are antisymmetric in their indices, we have
$\chi^{\lambda\nu\sigma\kappa}=-\chi^{\lambda\nu\kappa\sigma}=
-\chi^{\nu\lambda\sigma\kappa}$.  An antisymmetric pair of indices
corresponds, in 4D, to six independent components. Thus, the
constitutive tensor can be considered as a $6\times 6$ matrix with 36
independent components, see (\ref{DEB}) and (\ref{HEB}).

A $6\times 6$ matrix can be decomposed in its tracefree symmetric part
(20 independent components), its antisymmetric part (15 components),
and its trace (1 component). On the level of
$\chi^{\lambda\nu\sigma\kappa}$, this {\it decomposition} is reflected
in \cite{Birkbook}
\begin{eqnarray}\label{dec}
  \chi^{\lambda\nu\sigma\kappa}&=&\,^{(1)}\chi^{\lambda\nu\sigma\kappa}+
  \,^{(2)}\chi^{\lambda\nu\sigma\kappa}+
  \,^{(3)}\chi^{\lambda\nu\sigma\kappa}\,.\\ \nonumber 36
  &=&\hspace{15pt} 20\hspace{15pt}\oplus \hspace{15pt}15\hspace{15pt}
  \oplus \hspace{25pt}1\,.
\end{eqnarray}
The third part, the {\it axion} part, is totally antisymmetric and as
such proportional to the Levi-Civita symbol, $
^{(3)}\chi^{\lambda\nu\sigma\kappa}:= \chi^{[\lambda\nu\sigma\kappa]}
=\widetilde{\a}\, \widetilde{\epsilon}^{\lambda\nu\sigma\kappa}$.
Here, the totally antisymmetric Levi-Civita symbol is $
\widetilde{\epsilon}_{\lambda\nu\sigma\kappa}=\pm 1,0\,$; we denote
pseudotensors with a tilde. The second part, the {\it skewon} part, is
defined according to $ ^{(2)}\chi^{\mu\nu\lambda\rho}:=\frac
12(\chi^{\mu\nu\lambda\rho}- \chi^{\lambda\rho\mu\nu})$.  If the
constitutive equation can be derived {}from a Lagrangian, which is the
case as long as only reversible processes are considered, then
$^{(2)}\chi^{\lambda\nu\sigma\kappa}=0$. We will assume this condition
henceforth. Below, for Cr$_2$O$_3$, it will be verified
experimentally. The {\it principal} part
$^{(1)}\chi^{\lambda\nu\sigma\kappa}$ has the symmetries
$^{(1)}\chi^{\lambda\nu \sigma\kappa}=
{}^{(1)}\chi^{\sigma\kappa\lambda\nu}$ and
$^{(1)}\chi^{[\lambda\nu\sigma\kappa]}=0$.  The tensor
$\chi^{\lambda\nu\sigma\kappa}$ has now $20+1$ independent components
and the constitutive law reads
\begin{equation}\label{constit7}
  { \frak{G}^{\lambda\nu}=\frac
    12\left({}^{(1)}{\chi}^{\lambda\nu\sigma\kappa} +\widetilde{\a}\,
      \widetilde{\epsilon}^{\lambda\nu\sigma\kappa}\right)F_{\sigma\kappa}\,.}
\end{equation}
We can express the axion piece $\widetilde{\a}$ directly in the
constitutive tensor (\ref{dec}). With
$^{(2)}\chi^{\lambda\nu\sigma\kappa}=0$, we find
\begin{eqnarray}\label{pss1}
  \widetilde{\a}=\frac{1}{4!}
  \widetilde{\epsilon}_{\lambda\nu\sigma\kappa} \chi^{\lambda\nu\sigma\kappa}
=\frac{1}{3}\left(\chi^{0123}+ \chi^{0231} +
    \chi^{0312}\right)\,.
\end{eqnarray}
This 4D pseudoscalar (or axion piece) of
$\chi^{\lambda\nu\sigma\kappa}$ will be determined for Cr$_2$O$_3$.

We split (\ref{constit7}) into space and time \cite{Birkbook}. Then we
recover equations of the form of (\ref{DEB}) and  (\ref{HEB}), but
with an exact relativistic meaning ($a,b,\dots=1,2,3$):
\begin{eqnarray}\label{explicit3}
  {D}^a\!&=\!& {\varepsilon^{{ab}}}\,E_b + {\gamma^a{}_b}\, {B}^b +
  {\widetilde{\a}}\,B^a \,,\\ {H}_a\!  &=\!  & { \mu_{ab}^{-1}} {B}^b - {
    \gamma^b{}_a}E_b - {\widetilde{\a}}\,E_a\,.\label{explicit4}
\end{eqnarray}
We have the 6 permittivities $\varepsilon^{ab}= \varepsilon^{ba}$, the
6 permeabilities $\mu_{ab}=\mu_{ba}$, and the 8+1 magnetoelectric
pieces $\g^a{}_b$ (its trace vanishes, $\g^c{}_c=0$) and
$\widetilde{\a}$, respectively. Equivalent constitutive relations were
formulated by Serdyukov et al.\ \cite{Serdyukov}, p.86, and studied in
quite some detail. It is remarkable, as can be recognized {}from
(\ref{constit7}) and (\ref{pss1}), that $\widetilde{\a}$ is a 4D
pseudoscalar.

\section{Chromium oxide C${\rm r}_2$O$_3$ and its constitutive
  law}

\begin{figure}\label{Fig1}
\includegraphics[width=8.5cm]{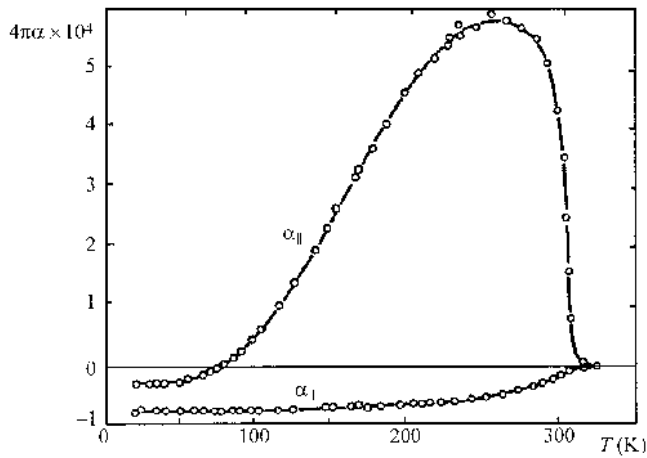}
\caption { The ME$_{\rm E}$ effect (linear magnetoelectric effect with
  electric field-induced magnetization) of Cr$_2$O$_3$: Temperature
  dependence of the magnetoelectric components $\a_{||}$ and $\a_\bot$
  according to Astrov \cite{Astrov}.}
\end{figure}

On the basis of neutron scattering data and susceptibility
measurements of the antiferromagnetic chromium sesquioxide
Cr$_2$O$_3$, Dzyaloshinskii \cite{Dz1} was able to establish the
magnetic symmetry class $\overline{3}{}'m'$ of a Cr$_2$O$_3$ crystal.
On this basis, he formulated the following constitutive law for
Cr$_2$O$_3$:
\begin{eqnarray}\label{DH1} D_{x,y} &=& \varepsilon_{\bot}
  \varepsilon_0 E_{x,y} +\frac{\alpha_{\bot}}{c}\,H_{x,y}\,,\\
  \label{DH3} D_z &=&
  \varepsilon_{||}\,\varepsilon_0 E_z +\frac{\alpha_{||}}{c}\, H_z\,,\\
  \label{BE1}
  B_{x,y} &=& \mu_{\bot}\mu_0 H_{x,y} +\frac{\alpha_{\bot}}{c}\,
  E_{x,y}\,,\\\label{BE3} B_z &=&
  \mu_{||}\,\mu_0 H_z +\frac{\alpha_{||}}{c}\, E_z\,.
\end{eqnarray}
The $z$-axis is parallel to the trigonal (and the optical) axis of the
crystal. The permittivities parallel and perpendicular to the $z$-axis
are denoted by $\varepsilon_{||},\varepsilon_{\bot}$, analogously the
permeabilities by $ \mu_{||}, \mu_{\bot}$, and the magnetoelectric
moduli by $\alpha_{||},\alpha_{\bot}$. Note that all these moduli are
dimensionless (in {\it all} systems of units).

The theory (\ref{DH1}) to (\ref{BE3}) and also the corresponding
measurements were made in the $(D,B)$ system. However, in order to get
to the manifestly relativistic covariant $(D,H)$ representation
(\ref{explicit3}),(\ref{explicit4}), we have to resolve the $(D,B)$
system with respect to $D$ and $H$ and to compare the corresponding
coefficients. We find magnetoelectric matrix
\begin{equation}\label{gamma}
  \gamma^a{}_b= \frac{Y_0}{3}\left(\frac{\a_\bot}{\mu_\bot}- 
\frac{\a_{||}}{\mu_{||}} \right)
  \begin{pmatrix}1&0&0\cr
    0&1&0\cr
    0&0&\hspace{-6pt}-2 \end{pmatrix}\,
\end{equation}
\begin{figure}[ht]
\includegraphics[width=260pt,height=200pt]{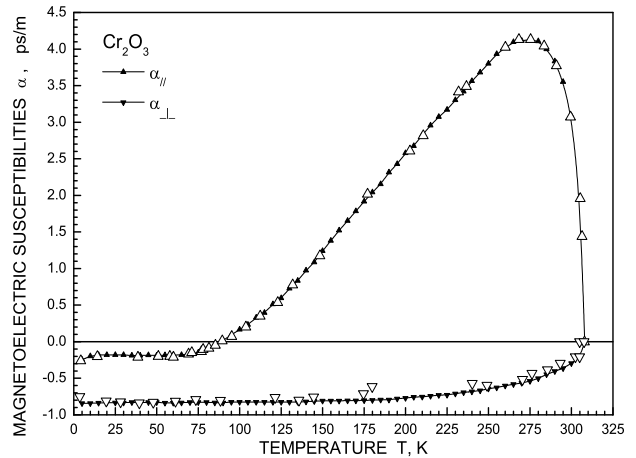}
\vspace{-30pt}
\caption{The ME$_{\rm H}$ effect (linear magnetoelectric effect with
  magnetic field-induced polarization) of Cr$_2$O$_3$: Temperature
  dependence of the magnetoelectric moduli $\a_{||}$ and $\a_\bot$
  according to Wiegelmann et al.\ \cite{Wiegelmann}, Fig.2.  Their
  measured values are denoted by large empty triangles, pointing up or
  down; superimposed are the interpolated and digitalized point, small
  black triangles, pointing up or down. The curve was determined by
  using B-splines. One should also compare Rivera \cite{Rivera1} and
  Wiegelmann \cite{WiegelmannDr}. The relative sign between $\a_{||}$
  and $\a_\bot$ was taken {}from Astrov \cite{Astrov}. The magnetic
  field $\mu_0H$ was below 6 {\it tesla}.}\label{Fig2}
\end{figure}
and the pseudoscalar or axion piece
\begin{equation}\label{axion}
  {  \widetilde{\a}= \frac{Y_0}{3}\left(2\,\frac{\a_\bot}{\mu_\bot}+ 
      \frac{\a_{||}}{\mu_{||}} \right)\,.}
\end{equation}
Eqs.(\ref{gamma}) and (\ref{axion}) can be collected in the
``relativistic'' $\a$-matrix
\begin{equation}\label{alpha}
^{\rm rel}\a^a{}_b:=\g^a{}_b+\widetilde{\a}\,\delta^a_b=Y_0\begin{pmatrix}
\frac{\a_\bot}{\mu_\bot}&0&0\\
0&\frac{\a_\bot}{\mu_\bot}&0\\
0&0&\frac{\a_{||}}{\mu_{||}}\end{pmatrix}\,.
\end{equation}

\section{Measurements on C${\rm r}_2$O$_3$}

Astrov \cite{Astrov}, see Fig.1, measured $\a_{||}$ and
$\a_\bot$ in an {\it electric} field $E$ according to (\ref{BE1}) and
(\ref{BE3}) and Rado \& Folen in a {\it magnetic} excitation $H$
according to (\ref{DH1}) and (\ref{DH3}).  Within the measurement
limits, they found the same values. This verifies that
$^{(2)}\chi^{\lambda\nu\sigma\kappa}=0$ and confirms Dzyaloshinskii's
theory (below the spin-flop phase). Later on, mostly measurements in
magnetic fields below $\mu_0H=6\,T$ were made, see Rivera
\cite{Rivera1} and Wiegelmann et al.\ \cite{Wiegelmann}.

Measurements of Wiegelmann et al.\ \cite{Wiegelmann} are plotted in
Fig.2. The maximum of $\a_{||}$ was found at about 275 $K$:
\begin{equation}\label{value2}
  \a_{||}\,({\rm at}\, 275\, K) \approx 4.13\, \frac{ps}{m}\times
  c\approx 1.238\times 10^{-3}\,.
\end{equation}
Now we can compute from the values of Fig.2 the pseudoscalar
(\ref{axion}). However, we need additionally the permeabilities
$\mu_{||}$ and $\mu_\bot$. They have been measured by Foner
\cite{Foner}. We find at 4.2 $K$, $\mu_\bot\approx 1.00147$ and the
maximum value near the N\'eel temperature $\approx 1.00162$. Since
$\mu_{||}$ deviates even less from $1$, we have $\mu\approx 1$. Then
(\ref{axion}) can be easily evaluated: $\widetilde{\a}\approx \frac
13\left(2\,{\a_\bot}+ {\a_{||}} \right)Y_0$. Our results are plotted
in Fig.3.

As we can see, for temperatures of up to about 163 $K$, the
pseudoscalar is negative, for higher temperatures positive until it
vanishes at the N\'eel temperature of about 308 $K$. For the maximum,
we find
\begin{eqnarray}\label{result1}
    \widetilde{\a}_{{\rm max}}\,({\rm at}\, 285\, K)
  \approx 3.10\times 
  10^{-4}\;Y_0 \stackrel{\rm SI}{\approx}& 0.822\;\frac{1}{M\Omega}\,.
\end{eqnarray}
We conclude that $\widetilde{\a}$ is fairly small but, for $T\ne 163
\,K$, definitely nonvanishing.

\begin{figure}[ht]
\includegraphics[width=260pt,height=200pt]{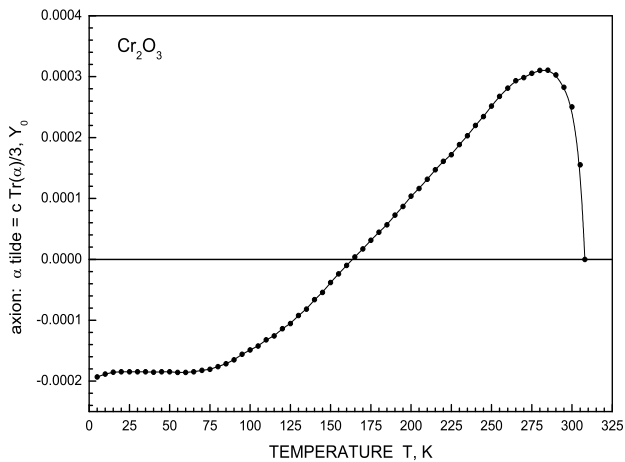}
\vspace{-10pt}
\caption{Our new result: The pseudoscalar or axion piece
  $\widetilde{\a}$ of the constitutive tensor
  $\chi^{\lambda\nu\sigma\kappa}$ of Cr$_2$O$_3$ in units of $Y_0$ as
  a function of the temperature $T$ in {\it kelvin}; here $Y_0$ is the
  vacuum admittance, which is 1 in Gaussian units and about 1/(377
  {\it ohm}) in SI. In the figure the tilde of $\,\widetilde{\a}$ is
  missing.}\label{Fig3}
\end{figure}

Does this result imply consequences also for the experimentalist?  We
think so for the following reason: As we saw above, $\mu_{||}$ as well
as $\mu_\bot$ are approximately one. Therefore the magnetoelectric
$\gamma$ matrix (\ref{gamma}) becomes
\begin{equation}\label{gamma'}
  \gamma^a{}_b\approx \frac{Y_0}{3}\left({\a_\bot}-{\a_{||}} \right) 
  \begin{pmatrix}1&0&0\cr
    0&1&0\cr
    0&0&\hspace{-6pt}-2 \end{pmatrix}\,.
\end{equation}
The question is now: {\it Can we find a substance in which} 
\begin{equation}\label{subst}
{\a_\bot}={\a_{||}}\,,
\end{equation}
that is, in which the matrix $\g^a{}_b$ vanishes {\em for all
  temperatures?} This challenge for experimentalists would be
interesting in the sense that then one would have a substance in which
the {\em only} magnetoelectric piece would be the pseudoscalar (or
axion) piece $\widetilde{\a}$. In other words, this substance would
display an {\em isotropic magnetoelectric effect.}

For a theoretician it could be of value if he/she looks for a
microscopic Hamiltonian. The pseudoscalar $\widetilde{\a}$ of the
magnetoelectric effect should have a different physical origin as
compared to the $\gamma$ part. Thus, it could be helpful for
developing microscopic models for the magnetoelectric effect.

\section{Properties of the pseudoscalar or axion piece}

Unlike the 3D vectors of the electric an the magnetic fields and the
3D tensors of permittivity $\ve^{ab}$, of permeability $\mu^{ab}$, and
of the magnetoelectric moduli $\gamma^a{}_b$, which all depend on the
choice of the reference frame and the local coordinates, the value of
$\widetilde{\a}$ is always the {\it same}. It is invariant under any
orientation-preserving transformation of frames and coordinates ---
and changes sign when the orientation is changed.

If we consider the pseudoscalar $\widetilde{\a}$ alone, then we can
take its constitutive law {}from (\ref{explicit3}) and
(\ref{explicit4}),
\begin{eqnarray}\label{Ax1}
  D^a=+\widetilde{\a}\,B^a\,,\quad
  H_a=-\widetilde{\a}\,E_a\,
\end{eqnarray}
or, in 4D, 
\begin{equation}\label{AX}\frak{G}^{\lambda\nu}=\widetilde{\a}\, 
  \widetilde{\epsilon}^
  {\lambda\nu\sigma\kappa}F_{\sigma\kappa}/2\,.
\end{equation} 
A space reflection
\begin{equation}\label{P}
D^a\rightarrow -D^a,\,H_a\rightarrow H_a,\,E_a\rightarrow
-E_a,\,B^a\rightarrow  B^a\,,
\end{equation}
as well and a time inversion 
\begin{equation}\label{T}
D^a\rightarrow D^a,\,H_a\rightarrow -H_a,\,E_a\rightarrow
E_a,\,B^a\rightarrow -B^a\,,
\end{equation}
see Janner \cite{Janner2} and Marmo et al.\ \cite{Bepe}, will turn
(\ref{Ax1}) into its negative,
\begin{eqnarray}\label{Ax1'}
  D^a=-\widetilde{\a}\,B^a\,,\quad
  H_a=+\widetilde{\a}\,E_a\,.
\end{eqnarray}
This is an expression of the {\it pseudo\/}scalar nature of
$\widetilde{\a}$. Therefore $\widetilde{\a}$ is P odd and T odd.

Moreover, the energy-momentum tensor for the electromagnetic field,
see \cite{Post,Birkbook},
\begin{equation}\label{em1}
  \frak{T}_\lambda{}^\nu=\frac 14\,\frak{G}^{\sigma\tau}
  F_{\sigma\tau}\d^\nu_\lambda
  -\frak{G}^{\nu\sigma}F_{\lambda\sigma}\,,
\end{equation}
if (\ref{AX}) is substituted, vanishes:
\begin{equation}\label{em3}
\frak{T}_\lambda{}^\nu\mbox{(of axion piece $\widetilde{\a}$)} =0\,.
\end{equation}
Thus, the electromagnetic energy density $\frak{T}_0{}^0$, the
energy-flux density (Poynting flux) $\frak{T}_0{}^b$ etc.\ of the
axion piece vanish.

\section{Analogues of the 4D pseudoscalar $\widetilde{\a}$ in
  network theory, in electrical engineering, and in particle physics}

The structure of the constitutive law (\ref{Ax1}) or (\ref{AX}) is not
unprecedented. In electrical engineering, in the theory linear
networks, more specifically in the theory of two ports (or four
poles), Tellegen \cite{Tellegen1948,Tellegen1956/7} came up with the
new structure of a {\it gyrator}, which is defined via
\begin{eqnarray}\label{gyrator}
v_1=-s\,i_2\,,\qquad
v_2=s\,i_1\,,
\end{eqnarray}
where $v$ are voltages and $i$ currents of the ports 1 and 2,
respectively. Let us quote {}from Tellegen \cite{Tellegen1956/7}, p.189:
``The ideal gyrator has the property of `gyrating' a current into a
voltage, and vice versa.  The coefficient $s$, which has the dimension
of a resistance, we call the gyration resistance; $1/s$ we call the
gyration conductance.'' The gyrator is a nonreciprocal network element.

If we turn to the electromagnetic field, then because of dimensional
reasons the quantities related to the {\it currents} $i_1,\,i_2$ are
the excitations $D^a,\,H_a$ and the quantities related to the {\it
  voltages} $v_1,\,v_2$ the field strengths $E_a,\,B^a$.  Then we find
straightforwardly the analogous relations
\begin{eqnarray}\label{gyrator*}
E_a=-s\,H_a\,,\qquad
B^a=s\,D^a\,.
\end{eqnarray}
If we rename the resistance $s$ according to $s=1/\widetilde{\a}$,
then (\ref{gyrator*}) and (\ref{Ax1}) coincide. Without the least
doubt, the gyrator is in the theory of electrical networks what the
axion piece is in magnetoelectricity. The axion piece `rotates' the
excitations, modulo an admittance, into the field strengths, as the
gyrator the currents into voltages.

These analogies or rather isomorphisms carry even further. In 2005,
Lindell \& Sihvola \cite{LindSihv2004a}, see also \cite{Ismobook},
introduced the new concept of a {\it perfect electromagnetic
  conductor} (PEMC). It obeys the constitutive law (\ref{Ax1}) or
(\ref{AX}). The PEMC is a generalization of the perfect electric and
the perfect magnetic conductor. In this sense, it is the `ideal'
electromagnetic conductor that can be hopefully built by means of a
suitable {\it metamaterial,} see Sihvola \cite{metaAri}.

Continuing with our search for isomorphisms, we turn to axion
electrodynamics, see Ni \cite{Ni} and Wilczek \cite{Wilczek87} and,
for more recent work, Itin \cite{Itin2004}, \cite{Itin2007}. If
for vacuum electrodynamics we add to the usual Maxwell-Lorentz
expression specified in (\ref{strel}) an axion piece patterned after
the last term in (\ref{constit7}), then we have the constitutive law
for axion electrodynamics,
\begin{equation}\label{axel}
  \frak{G}^{\lambda\nu}=Y_0\,\sqrt{-g}F^{\lambda\nu}+\frac
  12\,\widetilde{\a}\, \widetilde{\epsilon}^
  {\lambda\nu\sigma\kappa}F_{\sigma\kappa}\,.
\end{equation}

In Cr$_2$O$_3$ we have $\widetilde{\a}\approx 10^{-4}Y_0$.
It is everybody's guess what it could be for the physical vacuum. In
elementary particle theory one adds in the corresponding Lagrangian
also kinetic terms of the axion \`a la $\sim g^{\mu\nu}
\partial_\mu\widetilde{\a}\,\partial_\nu\widetilde{\a}$ and possibly a
massive term $\sim m_{\widetilde{\a}}^2\,\widetilde{\a}^2$.  However,
this hypothetical $P$ odd and $T$ odd particle has not been found so
far, in spite of considerable experimental efforts, see Davis et al.
\cite{Davis2007}.

The axion shares its $P$ odd and $T$ odd properties with the
$\widetilde{\a}$ piece of Cr$_2$O$_3$, with the gyrator, and with the
PEMC. One may speculate whether an axion detector made of Cr$_2$O$_3$
crystals could enhance the probability of finding axions.

\section{Conclusions} 

Our results can be summed up as follows:
  
$\bullet$ The magnetoelectric pseudoscalar of Cr$_2$O$_3$ is
temperature dependent and of the order of $10^{-4}Y_0$. Thus, the Post
constraint is invalid in general.

$\bullet$ Our result and the formalism is of general interest to areas
such as multiferroics and similar materials where there is a need for
a generalized and concise description of the phenomena.

$\bullet$ We suggest to the experimentalist to search for a substance
with an {\em isotropic\/} magnetoelectric effect.

$\bullet$ Since the gyrator and the PEMC are established notions in
network theory and in electrical engineering, these two notions,
together with the isomorphic structure of the pseudoscalar piece in
Cr$_2$O$_3$, support the existence of a fundamental axion field
coupled to conventional vacuum electrodynamics. Thus, axion
electrodynamics gains plausibility by our results.  \bigskip

\noindent{\it Acknowledgments}. One of us (F.W.H.) is very grateful
for useful discussions with Yakov Itin (Jerusalem), Ari Sihvola
(Helsinki) and with M.~Braden, T.~Nattermann and A.~Rosch (all from
Cologne).  Financial support from the DFG (HE 528/21-1) is gratefully
acknowledged.  

\end{document}